# A Compact, High Performance Atomic Magnetometer for Biomedical Applications

**Vishal K. Shah[1] and Ronald T. Wakai[2]**

[1]QuSpin Inc., Westminster, CO, 80021

[2]University of Wisconsin, Madison, WI, 53705

EMAIL: vshah@quspin.com

**Abstract**

We present a highly sensitive room-temperature atomic magnetometer (AM), designed for use in biomedical applications. The magnetometer sensor head is only 2x2x5 cm$^3$ and it is constructed using readily available, low-cost optical components. The magnetic field resolution of the AM is <10 fT/√Hz, which is comparable to cryogenically cooled superconducting quantum interference device (SQUID) magnetometers. We present side-by-side comparisons between our AM and a SQUID magnetometer, and show that equally high quality magnetoencephalography (MEG) and magnetocardiography (MCG) recordings can be obtained using our AM.

**1. Introduction**

Biomagnetism involves measurement of extremely weak magnetic fields originating from biological systems, including the human body. The most important and most extensively investigated biomagnetism signals are the magnetoencephalogram (MEG) and the magnetocardiogram (MCG), which are the magnetic analogs of the EEG and ECG, respectively. The study of biomagnetism was enabled by the advent of the superconducting quantum interference device (SQUID) magnetometer in the 1960s [1], and SQUIDs are still the most sensitive commercially available magnetic field detectors. In recent years, however, atomic magnetometer (AM) technology has advanced significantly, and laboratory prototypes with sensitivity exceeding that of SQUID magnetometers have been demonstrated [2]. A major advantage of the AM over SQUID magnetometers is that the AM does not require cryogenic cooling. By eliminating the need for complex, expensive cryogenic equipment, AMs can drastically reduce the cost of MEG/MCG instrumentation.

AMs based on alkali atoms enclosed within a vapor cell were first developed in the 1950's [3], [4]. In 1969, Dupont-Roc and coworkers developed a zero-field version of this AM with nearly 10 fT/√Hz level sensitivity [5]. In 1973, Tang and coworkers discovered a phenomenon, which led to suppression of spin-exchange relaxation in alkali atoms, opening a pathway to miniaturization of highly sensitive AMs operating in a low magnetic field [6]. In 2003, Romalis and coworkers used this discovery to demonstrate an ultra-sensitive AM with sub-femtotesla level sensitivity [7]. The AMs operating in this regime are now referred to as Spin-Exchange Relaxation-Free (SERF) magnetometers. In 2007, Shah and coworkers developed a compact version of the SERF AM, using a millimeter-scale microfabricated vapor cell and the simplified detection scheme developed by Dupont-Roc and coworkers [8]. Recently, a fully integrated version of the SERF chip-scale atomic magnetometer (CSAM) was developed at the National Institute of Standards and Technology (NIST) [9].

Here we describe a low-cost, compact AM that is a viable alternative to a SQUID magnetometer for biomagnetism applications requiring compact sensors with high sensitivity. Until now, AMs either lacked sufficient sensitivity or were too large and complex for clinical applications. The AMs described here have size and sensitivity similar to that of SQUID magnetometers, and they are manufactured using commercial off-the-shelf components with simple assembly techniques. They also have a precisely defined sensitive axis and can be integrated into a large, dense array for MEG source localization applications; thus, they are suitable as a low-cost, drop-in replacement for SQUID magnetometers in biomagnetism instrumentation.

The purpose of the study was to characterize the sensitivity and bandwidth of our AM and to compare its performance with that of a commercial SQUID biomagnetometer by making MCG and MEG recordings in the same subjects. Similar comparisons have also been reported in past [10], [11], [12], [13].

## 2. Methods

Design of Atomic Magnetometer

The overall design and operation of our AM is similar to that of the earlier NIST CSAM that used a single optical beam. A detailed discussion of the methods can be found in Ref. [14]; only a brief summary is given here. The magnetometer consists of three main components: (i) a resonant light source, (ii) a transparent glass alkali ($^{87}$Rb) vapor cell, and (iii) a photodiode to monitor the intensity of the light transmitted through the vapor cell. The resonant light is generated using a narrow linewidth laser with its optical wavelength tuned to the D1 transition of $^{87}$Rb atoms. The resonant light is used to spin-polarize the alkali atoms, and a zero-field cross-over resonance is observed by sweeping the magnetic field about zero in a direction perpendicular to the optical beam. A maximum in the intensity of light transmitted through the vapor cell is seen when the magnetic field is precisely equal to zero. The zero-field resonance is a Lorentzian function of the magnetic field with a full-width at half maximum (FWHM) of about 30 nT. The magnetometer is locked to the peak of the zero-field resonance by applying a modulation field generated from an external magnetic coil and using feedback from a lock-in amplifier.

Figure 1a shows a picture of our prototype AM with integrated optics, vapor-cell, and photo-detector. Instead of using chip-scale components, the AM here is constructed using slightly larger but lower cost off-the-shelf optical components. The 1-mm microfabricated vapor cell used in the CSAM is replaced by a 4-mm Pyrex vapor cell. The narrower resonance and stronger signals from a larger vapor cell in our AM enables the magnetometer to reliably achieve SQUID-level performance without painstaking optimization. The magnetometer housing is made using a 3-D printer with high-temperature plastics. The housing has internal features that simplify the assembly process by allowing the optical components to be dropped and glued in place without manual adjustment. As shown in Figure 1a, the outside dimensions of the magnetometer are only 1.5x1.5x3 cm$^3$. With the thermally insulating protective jacket, the outside dimensions are 2x2x5 cm$^3$ (see Figure 1b). A set of three-axis magnetic coils is wrapped around the jacket to locally cancel residual magnetic fields. This eliminates a need for large external magnetic field and gradient compensation coils, allowing the magnetometer to be freely oriented in the shielded environment. The vapor cell in the magnetometer is heated to over 150 °C to achieve sufficient alkali density to suppress spin-exchange relaxation in a low field environment. With an efficient passive design to thermally isolate the vapor cell from the magnetometer housing, the outside surface temperature of the magnetometer rises to no more than two degrees above ambient temperature; thus, the magnetometer can be placed in close proximity or even in direct contact with the subject. The average distance between the outer surface of the AM and the center of the vapor cell is less than one cm. This distance is substantially shorter than the typical 2 cm distance between the SQUID coil and the outside of the SQUID dewar. The resonant light for optical pumping is delivered to the magnetometer using a polarization maintaining fiber. Using optical fiber splitters, light from a single-laser system can be distributed to multiple AM channels. The magnetic field resolution was measured in a three-layer, 20 cm inner diameter cylindrical μ-metal magnetic shield.

MCG and MEG Measurements

The laboratory testing and validation of the AMs was carried out at the University of Wisconsin Biomagnetism Laboratory. MCG and auditory evoked response (AER) MEG recordings were made. For the MCG study, the subjects were ten healthy adult subjects: 5 male and 5 female. All 10 subjects were studied during the same 2-hour session. For the AER study, the subjects were 4 healthy adult subjects: 3 male and 1 female. They were studied during the same 3-hour session. The protocols were approved by the institutional review board, and informed consent was obtained from all subjects. The measurements were performed in a two-shell magnetically shielded room (ETS-Lindgren, Glendale Heights, IL) with a dc shielding factor of about 500. The performance of the AMs was compared side-by-side with that of a commercial 7-vector channel SQUID magnetometer system (621 Biomagnetometer, Tristan Technologies, San Diego, CA), which records the x, y, and z components of the magnetic field at 7 locations. The center channel is surrounded by 6 others, equally spaced on a circle with 40 cm spacing between adjacent coils. The magnetic field resolution of each channel was approximately 5 fT/(Hz)$^{1/2}$. One important difference between a SQUID magnetometer and an AM is that the SQUID magnetometer's pickup coil can be wound to configure the device as a gradiometer, which greatly improves rejection of environmental interference. The channels of the Tristan 621 Biomagnetometer were configured as first-order gradiometers with 8-cm baseline. Using AMs, however, gradiometers are much more difficult to implement. Instead, reference channel noise cancellation was performed by using the first AM (AM 2) as a signal channel to record MEG activity and the second (AM 1) as a reference channel to monitor the ambient magnetic interference.

The MCG subjects lay supine on the patient table and the AM was positioned over the left ventricle at a location where the MCG signal was expected to have largest amplitude (*Figure 2a*). The MCG was recorded for 1 minute using a 1-100 Hz passband and 1 kHz sampling rate. Next, the SQUID magnetometer was centered at the same location, and the MCG was recorded using the same data acquisition parameters. The AM showed substantial environmental interference at frequencies less than 2 Hz and at several frequencies above 50 Hz, including 60 Hz. A 3-43 Hz digital filter was applied to remove the interference. The SQUID recordings were also filtered using the same 3-43 Hz filter, although they contained much less interference since they were configured as gradiometers. Averaged MCG waveforms were computed by choosing a representative QRS complex as a template and computing the cross-correlation between the template and the MCG signal. The peaks in the cross-correlation were used to time-align the beats. All the beats within a 30 s interval were used in computing the averaged waveforms. The PR, QRS, and QT intervals were measured.

The MEG subjects lay on their side and the AM was positioned over the parietal region, near an extremum of the AER (Figure 2b). The stimuli consisted of 150 tone bursts of frequency 1 kHz, duration 50 ms, and intensity in air of 60 dB, presented with a pseudorandom interstimulus interval of 1-3 s. The tone bursts were produced by a piezoelectric speaker placed in a small opening of the shielded room. The MEG was recorded continuously, along with a trigger, consisting of square pulses that were synchronized with the stimulus presentation. Next, the SQUID magnetometer was centered at the same location, and the MEG was recorded using the same stimulation and data acquisition parameters. Reference channel noise cancellation, as described above, was required to remove large interference from a nearby air handling fan, which was turned off for the MCG study but could not be turned off for the AER study. The recording was further band limited using a 3-48 Hz bandpass digital filter. The SQUID recordings were also filtered using the same digital filter. Averaged AERs were computed, using the trigger channel to time-align the 150 AERs. In addition to recording AERs, in one subject the spontaneous MEG was recorded with the AM positioned over the occipital cortex. The subject was instructed to open and close his eyes in order to demonstrate blocking of the alpha rhythm.

## 3. Results

Figure 1c shows the sensitivity of two of the AM prototypes placed adjacent to each other. The sensitivity of each of the two magnetometers was below 10 fT/√Hz and could be reliably reproduced. The magnetic Johnson noise from the shield, however, was approximately 15 fT/√Hz, which limited the measured sensitivity. By using the standard technique of forming a gradiometer by subtracting the common mode signal, we were able to recover the true sensitivity of the AM, which was approximately 6 fT/√Hz. The 3-dB bandwidth of the magnetometers was roughly 100 Hz, which is sufficient for most MEG and MCG applications. The theoretical photon shot noise-limited sensitivity of the AMs was roughly 1 fT/√Hz.

Figure 3a shows a side-by-side comparison of averaged MCG waveforms from an adult subject, acquired successively with the AM and the SQUID magnetometer. The SQUID data is from a channel with large amplitude that most closely resembled the AM data. The waveforms registered by the two systems were very similar in overall quality. They are also remarkably similar in morphology, given that the position and orientation of the magnetometers were only approximately the same and that the AM and the SQUID magnetometer measure the magnetic field and the magnetic field gradient, respectively. The inset in Figure 3a shows the raw, full bandwidth MCG trace after application of only a 60 Hz notch filter. Equally high quality signals were obtained from all ten subjects. In Figure 4, similar side-by-side comparisons of time averaged MCG recordings are shown for the other 9 subjects. The waveform time intervals (Table 1) obtained with the two systems were also highly consistent. The limits of agreement were ≤5 ms for QRS interval, ≤10 ms PR interval and ≤25 ms for QTc interval. These limits are small enough that the differences in the measurements are not clinically significant.

Figure 3b shows a 8-12 Hz filtered MEG recording that depicts the phenomenon of alpha blocking, in which the rhythm is diminished when the subject opens his eyes.

Figure 5 shows a comparison of AERs recorded using the SQUID magnetometer and AM in four subjects. The left graph for each subject in Figure 5 shows a superimposition, or "butterfly", plot of the SQUID channels. The right graph shows a comparison between the response from the AM and the response from a representative SQUID channel with similar morphology. Again, the AM and SQUID recordings are very similar in quality and appearance.

## 4. Discussions

The data presented here validate the excellent performance of our AM in comparison to a commercial SQUID magnetometer. Both the MCG and MEG data obtained with our AM were remarkably similar in morphology and quality to data obtained with a SQUID magnetometer. In comparison to the MEG studies of Johnson and co-workers and Romalis and co-workers, which averaged 320 [12] and 600 [15] AER trials, respectively, we were able to achieve a high signal-to-noise ratio by averaging just 150 trials, which is typical for an MEG study.

The magnetic field resolution of our AM was <10 fT/(Hz)$^{1/2}$, whereas commercial SQUID magnetometers can achieve a resolution of about 5 fT/(Hz)$^{1/2}$. In practice, however, the resolution is often limited by the noise within the shielded room. If required, higher AM sensitivity should be achievable with improvements in laser stabilization.

AMs offer several crucial advantages over SQUID magnetometers, the most important of which is low cost. AM-based MCG/MEG systems are expected to be far less expensive than SQUID-based systems because AMs do not require cryogenic cooling and can be manufactured using off-the-shelf commercial components. Another potential advantage of an AM system is its small overall size. Because a bulky cryogenic dewar is not needed, it should be possible to make measurements in compact human-sized cylindrical shields that cost nearly an order of magnitude less than a typical magnetically shielded room.

*Table 1:* MCG Waveform time interval (ms) comparison (SQUID/AM)

| # | RR | | PR | | P | | QRS | | QT | | QTc | |
|---|-----|------|-----|-----|----|----|----|----|-----|-----|-----|-----|
| 1 | 667 | 660  | 83  | 81  | 52 | 51 | 75 | 78 | 342 | 340 | 419 | 419 |
| 2 | 842 | 843  | 147 | 149 | 49 | 46 | 88 | 91 | 391 | 391 | 426 | 426 |
| 3 | 980 | 998  | 131 | 130 | 55 | 53 | 79 | 82 | 397 | 400 | 401 | 400 |
| 4 | 756 | 754  | 134 | 138 | 53 | 52 | 99 | 97 | 330 | 330 | 380 | 380 |
| 5 | 905 | 905  | 118 | 119 | 52 | 50 | 89 | 90 | 369 | 353 | 389 | 393 |
| 6 | 705 | 744  | 98  | 95  | 58 | 55 | 85 | 93 | 354 | 306 | 422 | 355 |
| 7 | 913 | 905  | 121 | 115 | 51 | 47 | 89 | 90 | 409 | 386 | 428 | 406 |
| 8 | 858 | 907  | 102 | 97  | 34 | 34 | 73 | 75 | 394 | 389 | 425 | 408 |
| 9 | 995 | 1120 | 124 | 122 | 49 | 44 | 84 | 72 | 417 | 404 | 418 | 382 |
| 10| 996 | 1000 | 101 | 90  | 49 | 41 | 73 | 67 | 375 | 372 | 376 | 372 |

Although we have demonstrated only single-channel MCG/MEG, we believe the AMs described here satisfy all the requirements for incorporation into multi-channel MCG and whole-head MEG systems: (i) The AMs described here are sufficiently compact for constructing MCG arrays and MEG helmets with high channel density. (ii) The sensitive axis of the AMs is along the long axis of the magnetometer. This allows the AMs to be closely spaced to measure the radial component of the magnetic field, which is the most important component. (iii) The position and orientation of each magnetometer channel can be precisely known, which is crucial for source localization and quantitative measurements. This is accomplished through the use of a modulation field produced by a magnetic coil wrapped on the jacket of the magnetometer, which precisely defines the sensitive axis of the magnetometer. (iv) Synthetic gradiometers can be formed to improve rejection of environmental interference by deploying three orthogonal AMs to serve as reference channels.

Currently, AMs cannot be used as direct replacements for SQUIDs for all applications because they have lower bandwidth and they lose sensitivity if operated outside a low field environment. Nonetheless, the AM demonstrated in this study fulfills the principal requirements that will allow its use for most biomagnetism applications. We believe it is only a matter of time before AM-based MCG/MEG systems are utilized for studies that are currently performed using SQUID-based systems.

**Acknowledgements**

This work was supported by the NIH grants R43 HD074495 and R01 HL063174.

**Figures**

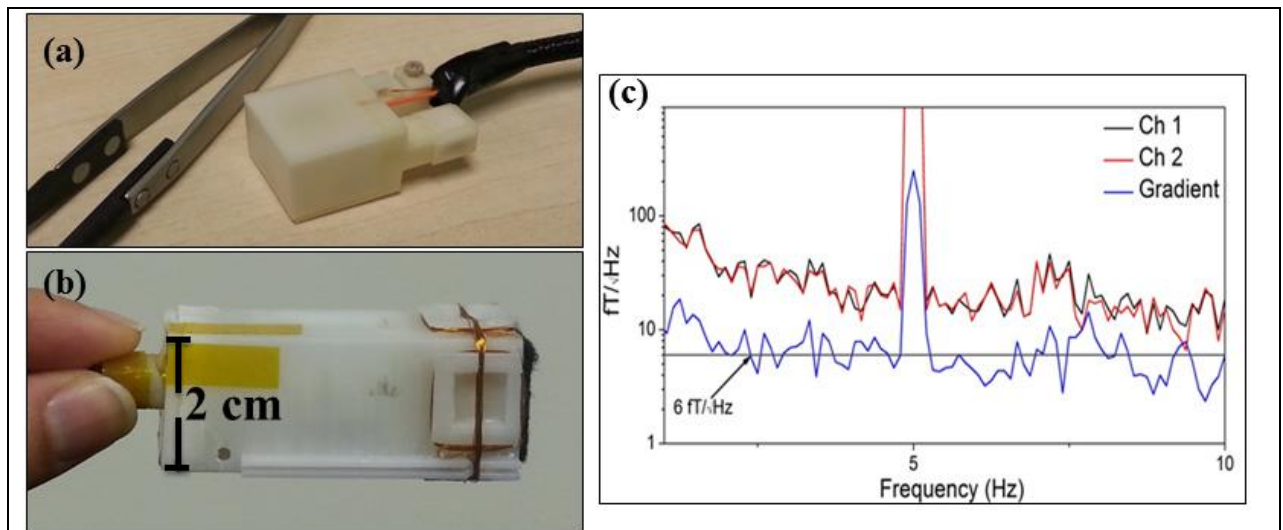

*Figure 1:* (*a*) Photograph of the prototype AM adjacent to optical tweezers. The external dimensions are 1.5x1.5x3 $cm^3$. (*b*) AM enclosed in an outer protective jacket (2x2x5 $cm^3$) with three-axis coils wrapped on the outside. (*c*) Magnetic field sensitivity of the AM prototypes. The red and the black traces are the magnetic noise power spectral density (PSD) of the two magnetometer channels measured within a magnetic shield, and the blue trace is the PSD of the difference of the outputs from the two magnetometers; i.e., synthetic gradiometer. The gradiometer arrangement cancels common mode magnetic Johnson noise from the magnetic shields, which allows the intrinsic sensitivity of the magnetometers to be deduced.

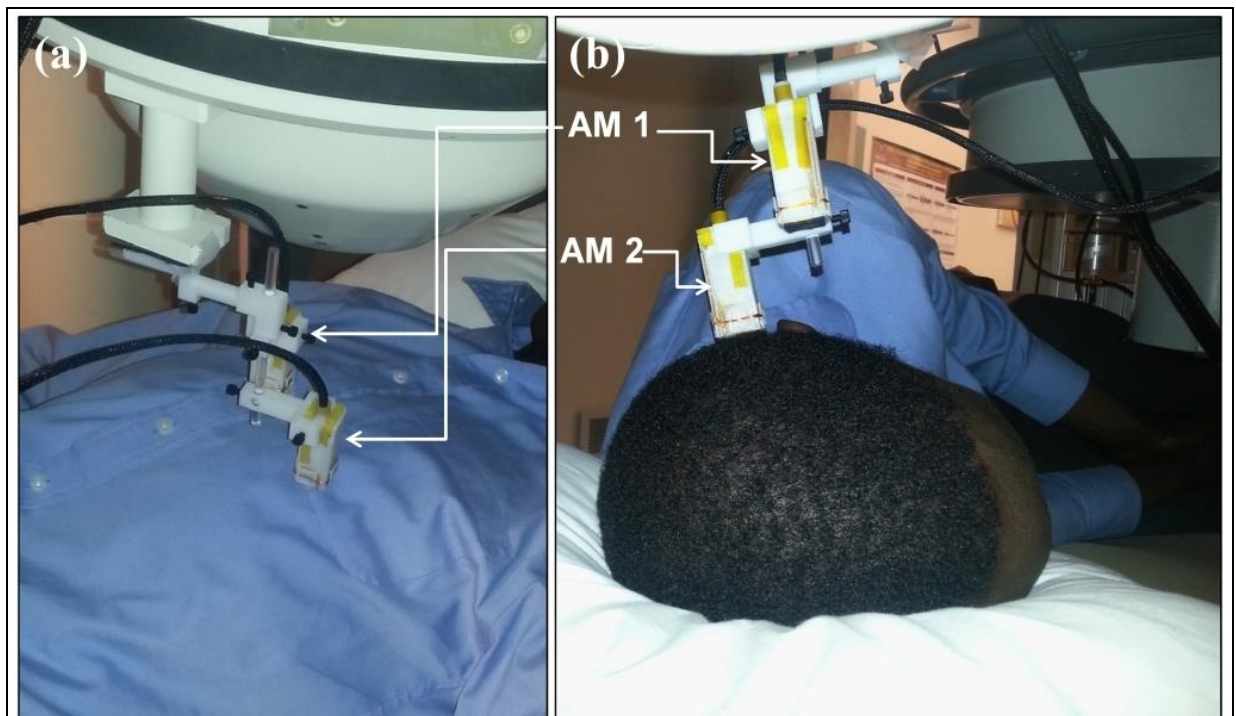

*Figure 2:* (*a*) A picture of two AMs positioned roughly over the chest of a subject for recording MCG. (*b*) A close-up picture of two AMs positioned over the parietal cortex in use for MEG-AER recordings. AM 2 which is closer to head was used for the actual measurements while AM 1 was used as reference sensor for background noise cancellation.

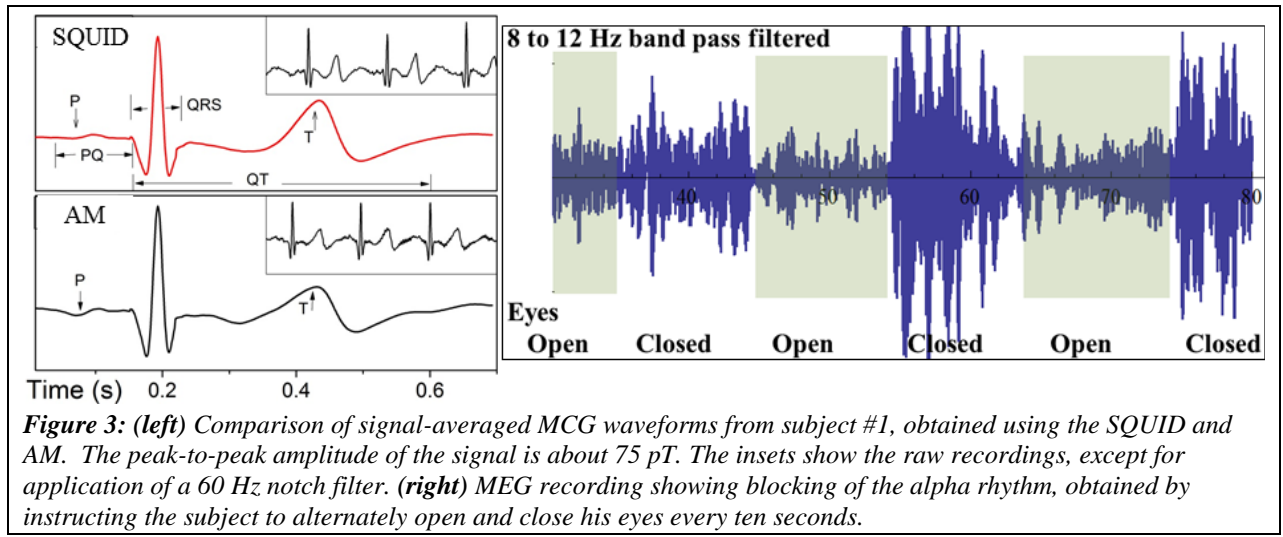

*Figure 3: (left)* Comparison of signal-averaged MCG waveforms from subject #1, obtained using the SQUID and AM. The peak-to-peak amplitude of the signal is about 75 pT. The insets show the raw recordings, except for application of a 60 Hz notch filter. *(right)* MEG recording showing blocking of the alpha rhythm, obtained by instructing the subject to alternately open and close his eyes every ten seconds.

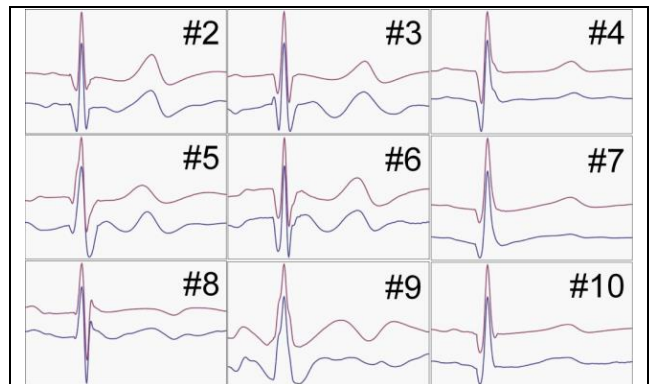

*Figure 4:* Thirty second time averaged MCG recordings of nine subjects. The red (upper) and the blue (lower) line show recording using SQUID and the AM respectively. Note slight differences in the waveform morphology due to mismatch in the exact location of the SQUID and the AM, and because the AM operated in magnetic field detection mode while the SQUID magnetometer was hard wired in gradiometer mode. The x-axis in each of the plots is time (0.7 s full-scale). The y-axis shows magnetometer output which is arbitrarily scaled for comparison.

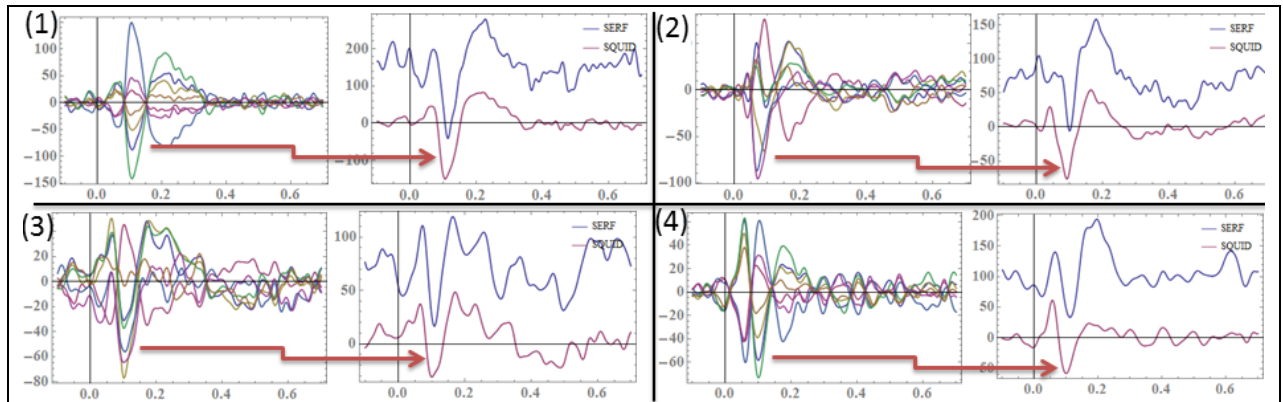

***Figure 5:*** *(1), (2), (3) and (4) show comparison Auditory evoked response recordings made using SQUID and AM of the four subjects. The left graph of each the four subjects is the AER recorded using a 7 channel SQUID system. (right) The response measured using AM (blue) and compared with the response from a SQUID channel (red) with similar morphology. The x-axis shows time in seconds and the y-axis is signal value in fT. The total measurement time was five minutes with approximately 165 averages of evoked responses separated by a randomized time interval between 1 and 3 seconds. The SQUID and AM recordings were arbitrarily offset on the y-axis of the plot for comparison.*